\documentclass[aps,prl,twocolumn,unsortedaddress,floatfix]{revtex4}
\usepackage{amsmath}
\usepackage{amsfonts}
\usepackage{amssymb}
\usepackage{natbib}
\usepackage{graphicx}
\usepackage{times}

\begin{document}

\title{Band filling and interband scattering effects in MgB$_2$: C vs Al doping}
\author{Jens Kortus}\email{j.kortus@fkf.mpg.de}
\altaffiliation[Current address: ]{Institut de Physique et de Chimie des 
Mat\'eriaux de Strasbourg, 23 Rue du Loess, F-67037 Strasbourg Cedex 2, France}
\author{Oleg V. Dolgov}
\author{Reinhard K. Kremer}%\email{rekre@fkf.mpg.de}

\affiliation{Max-Planck-Institut f{\"u}r Festk{\"o}rperforschung,
Heisenbergstr. 1, D-70569 Stuttgart, Germany}

\author{Alexander A. Golubov}
\affiliation{
MESA+ Research Institute and Faculty of Science and Technology, 
University of Twente, 7500 AE Enschede, The Netherlands
}

\date{\today}

\begin{abstract}
We argue, based on band structure calculations and Eliashberg
theory, that the observed decrease of $T_c$ of  Al and C doped MgB$_2$ samples
can be understood mainly in terms of a band filling effect due to the 
electron doping by Al and C. A simple scaling of the electron-phonon 
coupling constant $\lambda$ by the variation of the density
of states as function of electron doping is sufficient to capture the
experimentally observed behavior. 
Further, we also explain the long standing open question of the
experimental observation of a nearly constant $\pi$ gap as function of
doping by a compensation of the effect of band filling and interband scattering.
Both effects together generate a nearly constant $\pi$ gap and shift the
merging point of both gaps to higher doping concentrations, resolving the
discrepancy between experiment and theoretical predictions based on 
interband scattering only.
\end{abstract}

 \maketitle

The high critical temperature of 40 K in the simple
binary compound MgB$_2$ (Ref.\ \onlinecite{Nagamatsu}) was an unexpected 
present of nature to the scientific community. 
Now, after a few years of intense experimental and theoretical research 
the main features of superconductivity in this material seem well 
understood as due to a phonon mediated mechanism with different coupling
strengths to the electronic $\sigma$- and $\pi$-bands \cite{Kortus,Liu,Canfield,PhysicaC},
which leads to  the appearance of two distinct superconducting gaps.

Historically, two-band superconductivity  is an old
topic which has been proposed already shortly after 
the formulation of the BCS theory. Suhl, Matthias
and Walker \cite{Suhl} suggested a model for superconductivity in transition 
metals considering overlapping $s$- and $d$-bands. 
At the same time, Moskalenko formulated an extension of BCS theory for multiple bands \cite{Moskalenko}. 
In the early 1960 there have been experimental claims for the
observation of two-band superconductivity in some transition metals like e.g.\
V, Nb and Ta \cite{Shen} and later, in the 1980, in oxygen
depleted SrTiO$_3$ \cite{Bednorz}. 

However until now, MgB$_2$ appears to be first system for which multi-band
superconductivity has independently been evidenced by several experimental
techniques: heat capacity, tunneling spectroscopy, Raman
spectroscopy, penetration depth measurements, ARPRES and the analysis of the
critical fields \cite{PhysicaC}. 
The appearance of multiple gaps had been predicted theoretically \cite{Liu}
based on the electronic structure of MgB$_2$ \cite{Kortus,Pickett,Kong}.
The Fermi surface consists of four sheets:  
Two cylindrical sheets corresponding to quasi two-dimensional
$\sigma$-bands and two tubular networks derived from the
more three dimensional $\pi$-bands \cite{Kortus}. The phonons,
in particular the optical bond-stretching phonon branch along 
$\Gamma$-A \cite{Kong}, couple about three times stronger to the holes 
at the top of the $\sigma$-band as compared to the $\pi$-band 
\cite{Liu,Kong,Bohnen,Kunc,Golubov,Choi}.
Using linear response theory it is possible to calculate from first
principles the electron-phonon coupling (Eliashberg function)
which is needed as input for Eliashberg theory.
The solution of the Eliashberg equations allows for the calculation
of the superconducting gaps or thermodynamical properties like specific
heat in good agreement with the experiments \cite{Golubov,Choi}.

As for any anisotropic order parameter,
scattering by non-magnetic impurities should have a pair breaking effect, 
just as magnetic impurities have in conventional superconductors. 
Interband impurity scattering should lead to a decrease of 
$T_{\mathrm{c}}$ and if strong enough to a single (averaged)
order parameter \cite{Schopohl,GolMaz,Golubov}.
The interband impurity scattering between the $\sigma$- and $\pi$-bands is 
exceptionally small \cite{mazinimp},
due to the particular electronic structure of MgB$_{2},$ 
so that in the superconducting state, the two gaps in the 
$\sigma$- and the $\pi$-bands are preserved even in 'dirty' samples
with a considerably reduced $T_{\mathrm{c}}$ and a broad range of normal state 
resistivities.

\begin{figure}
\includegraphics[width=\linewidth,clip=true]{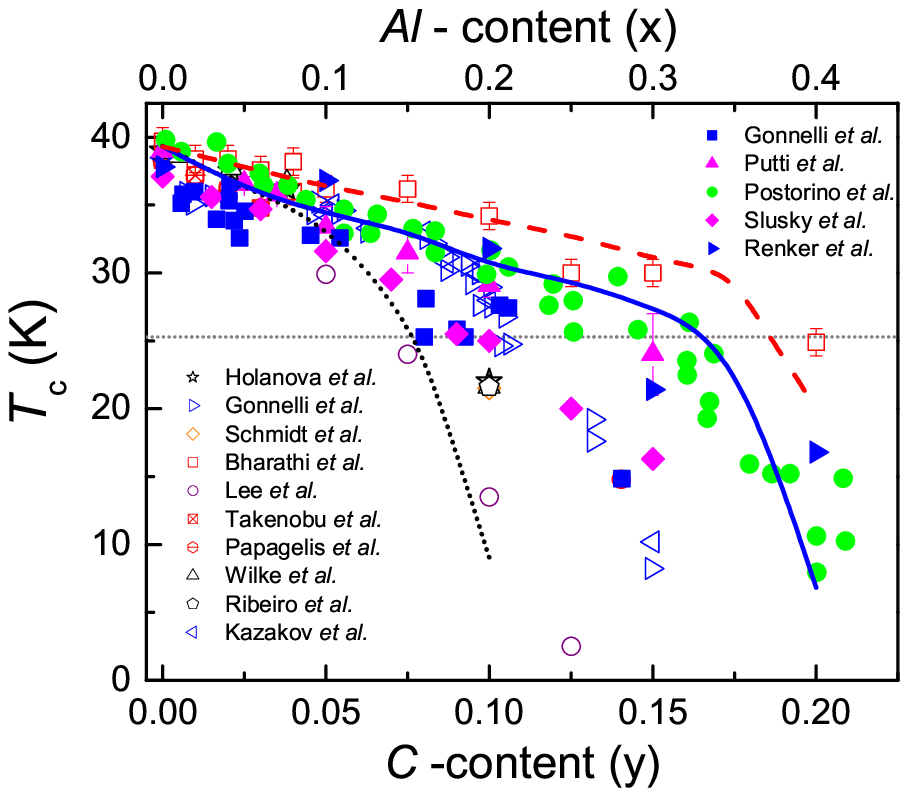}
\caption{\label{fig:tc-dop} Critical temperature $T_c$ as function of Al 
(filled symbols) and C (open symbols)versus doping concentration defined as
Mg$_{1-x}$Al$_x$(B$_{1-y}$C$_y$)$_2$.
Al-doping: ($\blacksquare$)\cite{Gonnelli1}, ($\blacktriangle$)\cite{Putti},
($\bullet$)\cite{Postorino},
($\blacklozenge$)\cite{Slusky}, ($\blacktriangleright$)\cite{Renker}
C-doping: ($\star$)\cite{Holanova}, ($\rhd$)\cite{Gonnelli1}, ($\diamondsuit$)\cite{Schmidt},
($\square$)\cite{Bharathi}, ($\bigcirc$)\cite{Lee}, ($\times$)\cite{Takenobu,Papagelis},
($\bigtriangleup$)\cite{Wilke}, ($+$)\cite{Ribeiro}, ($\lhd$)\cite{Kazakov}.
The lines present estimates based on Eliashberg theory with different
levels of approximation: dotted line with DOS from rigid band model, 
dashed line with DOS from VCA and solid line with DOS and phonon
renormalization from VCA. The horizontal dotted line is the lower limit
for interband scattering only.}
\end{figure}

The decrease of $T_c$ has been experimentally demonstrated 
by a series of substitution experiments in which Mg
has been replaced by Al and B by C \cite%
{Agrestini,Xiang,Li,Margadonna,Papavass,Bianconi,Pena,Postorino,
Castro,Putti,Ribeiro,Schmidt,Papagelis}. Similarly, irradiation with
neutrons leads also to a decrease of $T_{\mathrm{c}}$ \cite{Wangirra}.
Figure \ref{fig:tc-dop} shows a compilation of experimental data for the
critical temperature $T_c$ versus Al and C doping concentration.

For the two superconducting gaps it has been observed that
the $\sigma$ gap decreases with decreasing $T_{\rm c}$ 
and approaches the intermediate coupling value of 
2\,$\Delta$\,/$k_{\rm B}$\,$T_{\rm c}$ at $T_{\rm c} \sim$ 25 K. 
In most experimental reports the $\pi$ gap is found to be independent
on the $T_c$ of the sample and to remain close to the value of
$\sim$ 2 meV seen for undoped samples \cite{note}.

There have been recent reports by
Gonnelli and coworkers \cite{Gonnelli1,Gonnelli2} 
which demonstrate a different behavior of the superconducting gaps 
depending on the type of dopant. 
For C doped single crystals with composition Mg(B$_{1-y}$C$_y$)$_2$
($y$\,$\leq$\,0.132) their point contact spectroscopy measurements
show a merging of the $\sigma$ and $\pi$ gaps for the first time
\cite{Gonnelli1,Gonnelli2}. On the other hand, Gonnelli {\it et al.} also 
find that the behavior of Al doping in single crystals 
with composition Mg$_{1-x}$Al$_x$B$_2$ ($x$\,$\leq$\,0.21) 
is quite different. Even samples with very low $T_{\rm c}$ of about 20\,K
still exhibit distinct gaps, at critical temperatures
for which theoretical calculations based on 
Eliashberg theory including interband scattering always
predict a single order parameter only \cite{tunneling}.
Therefore, at present, there is disagreement between experiment and theory.

In the following we will argue that one essential ingredient to understand 
the behavior of $T_c$ is the effect of band filling of holes
in the $\sigma$-band due to electron doping. To understand the different
behavior of the two gaps in Al and C doped samples one additionally needs 
to consider interband scattering. While band filling will decrease
the superconducting gaps, interband scattering will
decrease the value of the larger gap and \textit{increase} the smaller one.
These two effects may compensate for the smaller $\pi$ gap and 
enable us to explain the observed nearly constant value of the small gap.

First, we will focus on the doping dependence of the critical temperature.
Figure \ref{fig:tc-dop} summarizes experimental results from different groups.
$T_c$ as function of Al and C doping shows very similar behavior
if the C doping is scaled by a factor of two as compared to the Al doping.
This follows naturally from the definition of the C doping concentration
per boron atom, as expressed in Mg$_{1-x}$Al$_x$(B$_{1-y}$C$_y$)$_2$,
with $x$ ($y$) for the amount of Al (C) doping.  
The importance of the band filling is already indicated by the
horizontal dotted line $\sim$25 K. This value would be the upper limit 
of $T_c$ due to the pair breaking effect of interband scattering only. 
If interband
scattering would be the only relevant mechanism no sample should show
a $T_c$ lower than indicated by the horizontal line. This is clearly not 
the case as shown in Fig.~\ref{fig:tc-dop}.

MgB$_2$ has a total of 0.26 holes: 0.15 holes in both $\sigma$ bands 
and the remaining 0.11 holes in the hole $\pi$-band. Al and C 
substitution will both dope electrons and therefore reduce the number of holes.
In a rigid band model the electron doping would be
defined with respect to the total number of holes in MgB$_2$ and 
simply corresponds to a shift of the Fermi level.
For small doping the $\sigma$-band DOS is practically constant as expected 
from the quasi two-dimensional character of the $\sigma$-bands. 
After adding 0.15 electrons the $\sigma$ bands become nearly filled and the 
DOS starts to decrease rapidly.
The coupling of the $\sigma$-holes to the optical bond-stretching $E_{2g}$ phonons
drives the superconductivity in this material and determines $T_c$.
Therefore, we just scale the band splitted electron-phonon 
Eliashberg functions $\alpha_{ij}^2F$ \cite{Golubov}
and the $\mu^{*}$-matrix \cite{comment} with the
the change of the $\sigma$- or $\pi$-band DOS as function of doping.
We use the Eliashberg functions for pure MgB$_2$ calculated from 
first-principles linear response theory \cite{Kong}, 
which have been used successfully to describe the specific heat \cite{Golubov}, 
tunneling \cite{tunneling} and penetration depth \cite{penetration}. 

The dotted line shown in Fig.~\ref{fig:tc-dop} corresponds to the rigid band scaling.
The decrease in $T_c$ for small doping concentrations is well reproduced
and originates from the small $k_z$ dispersion of the $\sigma$ bands along the $\Gamma$-A
line. The $\sigma$-band Fermi surfaces are not perfect cylinders but are slightly
warped (see Fig.\ 3, Ref.~\cite{Kortus}). For larger doping concentration, 
$T_c$ obtained from this simple model decreases faster than observed in
experiment. 
This is not surprising because we used the unperturbed band structure of pure 
MgB$_2$ not taking into account neither alterations of the bands due to doping 
nor the change of the phonon frequencies.

To correct for this failure we further calculate the change of the DOS using
the virtual crystal approximation (VCA). In order to simulate the
doping of  Al, we replace the Mg atom with a virtual atom with charge
$Z= x Z_{Al} + (1-x) Z_{Mg}$ and recalculate the electronic band structure
self-consistently using the full potential LMTO method \cite{Savrasov}. 
In agreement with \cite{Massidda}, we find a slower decrease of the 
$\sigma$-band DOS. Using the DOS from the VCA to scale 
$\alpha_{ij}^2F$ we solve the Eliashberg equations and obtain 
a slower decrease of $T_c$ (dashed line in Fig.~\ref{fig:tc-dop})
in better agreement with the experimental observations. 
Recent supercell calculations indicate an even slower $\sigma$-band 
filling\cite{Pick-Cdop} compared to the VCA.

An additional effect of doping will be the hardening of the $E_{2g}$ phonon
branch which will decrease the electron-phonon coupling $\lambda \sim 1/\omega^2$ 
\cite{Pickett}.
In order to take this effect into account we also calculated the
$E_{2g}$-$\Gamma$-point frequency in the VCA using linear response methods \cite{Savrasov}.
The final result from scaling $\alpha_{ij}^2F$ by the DOS and the 
$E_{2g}$ phonon frequency is shown by the solid line in Fig.~\ref{fig:tc-dop}. 
The agreement with experiment improved significantly.

Band filling with the corresponding changes in the DOS seems to be sufficient
to understand the behavior of $T_c$ as function of doping. However, this
is not sufficient to understand the evolution of the superconducting gaps,
because there should be no difference in the behavior 
between Al and C doping because both are electron dopants. 
We now argue that the additional ingredient to understand this behavior 
is interband scattering.

\begin{figure}
\includegraphics[width=\linewidth,clip=true]{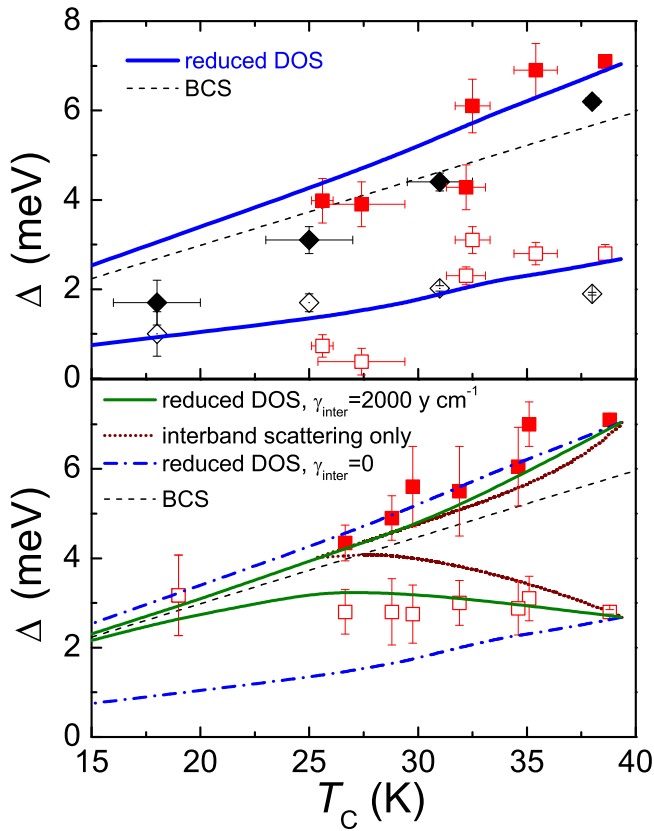}
\caption{\label{fig:gaps-tc}%
Upper panel Al doping: 
Superconducting $\sigma$-gap (upper curve and filled symbols) and
$\pi$-gap (lower curve and open symbols) as function of 
critical temperature $T_c$ obtained from
the solution of the Eliashberg equations with scaled $\alpha_{ij}^2F$ 
without interband scattering (solid lines) compared to experimental results 
($\square$)  \protect\cite{Gonnelli1} and
($\diamond$) \protect\cite{Putti}. 
Lower panel C doping: 
The solid lines show the solution of the Eliashberg equations 
with interband scattering rate $2000\cdot y$ cm$^{-1}$ 
compared to the experimental results ($\square$) \protect\cite{Gonnelli1,Gonnelli2}.
The limiting cases of interband scattering only (dotted lines) 
or scaling of $\alpha_{ij}^2F$ (dash-dotted lines) are also shown.
The dashed straight line indicates the BCS gap relation 
for $\Delta$/$T_{\rm c}$.}
\end{figure}

In the upper panel of Fig.~\ref{fig:gaps-tc} we plot the experimental 
$\sigma$- and $\pi$-gaps for Al doped crystals as
obtained by Gonnelli \textit{et al.} \cite{Gonnelli1}
and by Putti \textit{et al.} \cite{Putti} as function 
of the critical temperature of the samples.
Together with the experimental data we display the results from 
the solution of the two-band Eliashberg equations 
without interband scattering but the Eliashberg functions  
scaled by the change of DOS and phonon frequency as described above. 

The agreement with experiment seems to be reasonable.
The results by Putti \textit{et al.} \cite{Putti} show 
a merging of the gaps for Al-doping, however both gaps
have a ratio of $\Delta / k_B T_c$ somewhat lower than the canonical 
BCS ratio as indicated by a dashed line in Fig.~\ref{fig:gaps-tc}, 
which casts doubt on this data point.
However, it is still premature to give a definite answer
because more experimental data for single crystals and
high doping concentrations will be required for a complete picture.
Based on the available data which show no clear merging of the two gaps, 
we conclude that the interband scattering in Al doped samples is
small, even at high doping concentrations.

The  experimental results for C doped single crystals indicate
for the first time a merging of the two superconducting gaps,
which is a clear manifestation of interband scattering. 
In the lower panel of Fig.~\ref{fig:gaps-tc} we show the experimental 
results \cite{Gonnelli1} together with our Eliashberg theory calculations. 
In difference to the previous case we include the interband scattering in our 
calculations, which will also cause an additional reduction of $T_c$. 
Using a simple linear relation of the interband scattering rate to the doping 
concentration ($\gamma_{\mbox{inter}}=2000\cdot y$ cm$^{-1}$)
we find e.g.\ for 10-15\% C concentration an additional lowering 
of $T_c$ of about 6 K.

The two limiting cases (interband scattering only, scaling of $\alpha_{ij}^2F$) 
are also shown in the lower panel in Fig.~\ref{fig:gaps-tc}. 
The decrease of the DOS causes a decrease of both gaps, as 
can be seen from the dash-dotted lines. In contrast the interband scattering will 
decrease the $\sigma$ gap and \textit{increase} the $\pi$ gap. Both effects
can compensate each other resulting in the solid line, 
which includes the effects of the scaled $\alpha_{ij}^2F$ and interband 
scattering.
This may explain the experimental observation of a nearly constant 
$\pi$-gap as function of doping, which has been a long standing open question.

The difference in the magnitude of the interband scattering for Al and C doping 
can be easily rationalized. The $\sigma$ bond-orbitals are located in the boron plane
and there is not much weight of the $\sigma$-bands in the Mg plane. 
The $\pi$-orbitals are also centered at the boron plane, but extend
further out towards the Mg plane.
For that reason impurities in the boron plane are more effective
interband scatterers \cite{mazinimp,MazErw}. Therefore interband scattering due to 
doped C atoms replacing B atoms is much more likely than for Al doping.

In summary, we have shown that the variation in $T_c$ of Al and C doped 
samples of MgB$_2$ can be understood mainly 
as due to a simple effect of band filling. Al and C are both
electron dopants which reduce the number of holes at the top of the 
$\sigma$ bands together with a reduction of the electronic DOS.
Further, we suggest that the nearly constant $\pi$ gap as function of
doping can be understood due to a compensation of band filling and interband scattering.
The compensation of these effects shifts the merging point of both gaps to 
higher doping concentrations and lower $T_c$, 
resolving the discrepancy between experiment and theoretical predictions based on 
interband scattering only.

\begin{acknowledgments}
We thank J.\ K\"ohler, and A. Simon  for sharing with us their
insight into the chemistry of the substitution experiments
and O.\ K.\ Andersen and O.\ Jepsen for many enlightening discussions
and H.\ U.\ Habermeier for useful comments.
We are indebted to M. Putti and R.S. Gonnelli for sending us their experimental data.

\end{acknowledgments}


\begin{thebibliography}{99}

\bibitem{Nagamatsu}J. Nagamatsu, N. Nakagawa, T. Muranaka, Y. Zenitani, and J. Akimitsu, Nature (London) \textbf{63}, 401 (2001).

\bibitem{PhysicaC} See e.g. the special issue on MgB$_2$ in Physica C \textbf{385} (2003).

\bibitem{Canfield} P.C. Canfield and G. W. Crabtree, Physics Today March, p. 34 (2003).

\bibitem{Liu}
A.Y. Liu, I.I. Mazin, and J. Kortus, Phys. Rev. Lett. \textbf{87}, 087005 (2001).

\bibitem{Kortus}
J. Kortus, I. I. Mazin, K. D. Belashchenko, V. P. Antropov, and L. L. Boyer, Phys. Rev. Lett. \textbf{86}, 4656 (2001).


\bibitem{Suhl}H. Suhl, B.T. Matthias, and L.R. Walker, Phys. Rev. Lett. \textbf{3}, 552 (1959).


\bibitem{Moskalenko}A. Moskalenko, M.E.Palistrant, and V.M.Vakalyuk, Sov. Phys. Uspekhi \textbf{161}, 155 (1991)


\bibitem{Shen}L. Y. L. Shen, N. M. Senozan, and N. E. Phillips, Phys. Rev. Lett. \textbf{14}, 1025 (1965);
R. Radebaugh and P. H. Keesom, Phys. Rev. \textbf{149}, 209 (1966).




\bibitem{Bednorz} G. Binnig, A. Baratoff, H.E. Hoenig, and J.G. Bednorz, Phys. Rev. Lett. \textbf{45}, 1352 (1980).



\bibitem{Pickett}J.M. An, W.E. Pickett, Phys. Rev. Lett. \textbf{86}, 4366 (2001).


\bibitem{Kong}Y. Kong, O. V. Dolgov, O. Jepsen, and O. K. Andersen, Phys. Rev. B \textbf{64}, 020501(R) (2001).



\bibitem{Bohnen}K. P. Bohnen, R. Heid, and B. Renker, Phys. Rev. Lett. \textbf{86}, 5771 (2001).


\bibitem{Kunc}K. Kunc, I. Loa, K. Syassen, R. K. Kremer, and K. Ahn  J. Phys: Cond. Matter \textbf{13}, 9945 (2001).


\bibitem{Golubov}
A. A. Golubov, J. Kortus, O. V. Dolgov, O. Jepsen, Y. Kong, O. K. Andersen, B. J. Gibson, K. Ahn, and R. K. Kremer,  J. Phys: Condens. Matter \textbf{14}, 1353 (2002).


\bibitem{Choi}
H.J.~Choi, D. Roundy, H. Sun, M.L. Cohen, and S.G. Louie, \prb \textbf{66}, 020513 (2002);
H.J.~Choi, D. Roundy, H. Sun, M.L. Cohen, and S.G. Louie, Nature (London) \textbf{418}, 758 (2002).

\bibitem{Schopohl}N. Schopohl and K. Scharnberg, Solid State Commun. \textbf{22}, 371 (1977);
C.C. Sung and V. K.Wong, J. Phys. Chem. Solids \textbf{28}, 1933 (1967).


\bibitem{GolMaz}
A. A. Golubov and I. I. Mazin, Phys. Rev. B \textbf{55}, 15146 (1997).

\bibitem{mazinimp}I.I.~Mazin, O.K.~Andersen, O.~Jepsen, O.V.~Dolgov, J.~Kortus, A.A.~Golubov, A.B. Kuz'menko, and D. van der Marel, Phys. Rev. Lett. \textbf{89}, 107002 (2002).

\bibitem{Agrestini}
S. Agrestini, D. Di Castro, M. Sansoni, N. L. Saini, A. Saccone,
S. De Negri, M. Giovannini, M. Colapietro, and A. Bianconi, J.
Phys.: Condens. Matter \textbf{13}, 11689 (2001).

\bibitem{Xiang}
J. Y. Xiang, D. N. Zheng, J. Q. Li, L. Li, L. Lang, H. Chen, C. Dong, G. C. Che, Z. A. Ren, H. H. Qi, Y. Tian, Y. M. Ni, and Z. X. Zhao, Phys. Rev. B \textbf{65}, 214536 (2002).

\bibitem{Li}
J. Q. Li, L. Li, F. M. Liu, C. Dong, J. Y. Ziang, and Z. X. Zhao, Phys. Rev. B \textbf{65}, 132505 (2002).

\bibitem{Margadonna}
S. Margadonna, K. Prassides, I. Arvanitidis, M. Pissas, G. Papavassiliou, and A. N. Fitch, 
Phys. Rev, B \textbf{66}, 014518 (2002).

\bibitem{Papavass}
G. Papavassiliou, M. Pissas, M. Karayanni, M. Fardis, S. Koutandos and K. Prassides, 
Phys. Rev, B \textbf{66}, 140514 (2002).

\bibitem{Bianconi}
A. Bianconi, S. Agrestini, D. Di Castro, G. Campi, Z. Zangari, N. L. Saini, A. Saccone, S. De Negri, M. Giovannini, G. Profeta, A. Continenza, G. Satta, S. Massidda, A. Cassetta, A. Pifferi, and M. Colapietro, Phys. Rev. B \textbf{65}, 174515 (2002).


\bibitem{Pena} O. de la Pena,  A. Aguayo, and R. de Coss, Phys. Rev. B \textbf{66}, 012511 (2002).

\bibitem{Postorino}
P. Postorino, A. Congeduti, P. Dore, A. Nucara, A. Bianconi, D. Di Castro, S. De Negri, and A. Saccone, Phys. Rev. B \textbf{65}, 020507 (2001).

\bibitem{Castro}
D. Di Castro, S. Agrestini, G. Campi, A. Cassetta, M. Colapietro, A. Congeduti, A. Continenza, S. De Negri, M. Giovannini, S. Massidda, M. Tardone, A. Pifferi, P. Postorino, G. Profeta, and A.
Saccone, Europhys. Lett. \textbf{58}, 278 (2002).


\bibitem{Putti}
M. Putti, M. Affronte, P. Manfrinetti, and A. Palenzona, Phys. Rev. B \textbf{68}, 094514 (2003).

\bibitem{Ribeiro}
R.A. Ribeiro, S.L. Bud'ko, C. Petrovic, and P.C. Canfield, Physica C \textbf{384}, 227 (2003); R.A. Ribeiro, S.L. Budko, C. Petrovic, and P.C. Canfield, Physica C \textbf{385}, 16 (2003).

\bibitem{Schmidt}
H. Schmidt, K. E. Gray, D. G. Hinks, J. F. Zasadzinski, M. Avdeev, J. D. Jorgensen, and J. C. Burlea, Phys. Rev. B \textbf{68}, 060508(R) (2003).

\bibitem{Papagelis}
K. Papagelis, J. Arvanitidis, K. Prassides, A. Schenck, T. Takenbou, and Y. Iwasa, Europhys. Lett. \textbf{61}, 254 (2003).

\bibitem{Wangirra}
Y. Wang, F. Bouquet, I. Sheikin, P. Toulemonde, B. Revaz, M. Eisterer, H. W. Weber, J. Hinderer, and A. Junod, J. Phys.: Condens. Matter \textbf{15}, 883 (2003).

\bibitem{Gonnelli1} 
R.S. Gonnelli, D. Daghero, G.A. Ummarino, A. Calzolari, V. Dellarocca, V.A. Stepanov, S.M. Kazakov, J. Jun, J. Karpinski,  \eprint{cond-mat/0407267}.

\bibitem{Slusky}
J. S. Slusky, N. Rogado, K. A. Regan, M. A. Hayward, P. Khalifah, T. He, K. Inumaru, S. M. Loureiro, M. K. Haas, H. W. Zandbergen, and  R. J. Cava, Nature \textbf{410}, 343 (2001).

\bibitem{Renker}
B. Renker, K.B. Bohnen, R. Heid, and D. Ernst, H. Schober and M. Koza, P. Adelmann, P. Schweiss, and T. Wolf, Phys. Rev. Lett. \textbf{88}, 067001 (2002).

\bibitem{Holanova}
Z. Ho$\breve{\rm l}$anov$\acute{\rm a}$, P. Szab$\acute{\rm o}$,
P. Samuely, R. H. T. Wilke, S. L. Bud'ko, and P. C. Canfield, 
Phys. Rev. B \textbf{70}, 064520 (2004).

\bibitem{Bharathi}
A. Bharathi, S. Jemima Balaselvi, S. Kalavathi, G. L. N. Reddy, V. Sankara Sastry, Y. Hariharan, T. S. Radhakrishnan, Physica C \textbf{370}, 211 (2002).


\bibitem{Lee}
S. Lee, T. Masui, A. Yamamoto, H. Uchiyama, and S. Tajima, Physica C \textbf{397}, 7 (2003).

\bibitem{Takenobu}
T. Takenobu, T. Ito, Dam Hieu Chi, K. Prassides, and Y. Iwasa, Phys. Rev. B \textbf{64}, 134513 (2001).

\bibitem{Wilke}
R. H. T. Wilke, S. L. Bud'ko, P. C. Canfield, and D. K. Finnemore, Raymond J. Suplinskas, and S.T. Hannahs, Phys. Rev. Lett. \textbf{92}, 217003 (2004).

\bibitem{Kazakov}
S. M. Kazakov, R. Puzniak, K. Rogacki, A. V. Mironov, N. D. Zhigadlo, J. Jun, Ch. Soltmann, B. Batlogg, and J. Karpinski, \eprint{cond-mat/0405060}.

\bibitem{note} We note that there are some experimental reports \cite{Papagelis} which deviate from this general trend as e.g. the gap values obtained from the $\mu^+$SR study of
carbon\,-\,doped MgB$_2$.    
\bibitem{Gonnelli2}
R.S. Gonnelli, D. Daghero, A. Calzolari, G.A. Ummarino, V. Dellarocca, V.A. Stepanov, S.M. Kazakov, J. Jun, J. Karpinski, \eprint{cond-mat/0407265}.

\bibitem{tunneling}
 A.\ Brinkman, A.\ A.\ Golubov, H.\ Rogalla, O.\ V.\ Dolgov,
J.\ Kortus, Y.\ Kong, O.\ Jepsen, O.\ K.\ Andersen, Phys. Rev. B \textbf{65}, 180517(R) (2002).


\bibitem{comment}
I.I.~Mazin, O.K.~Andersen, O.~Jepsen, A.A.~Golubov, O.V.~Dolgov, and J.~Kortus, Phys. Rev. B \textbf{69}, 056501 (2004).


\bibitem{penetration}A.\ A.\ Golubov, A.\ Brinkman, O.\ V.\ Dolgov, J.\ Kortus, and O.\ Jepsen, Phys. Rev. B \textbf{66}, 054524 (2002).

\bibitem{Savrasov} S.\ Y.\ Savrasov, Phys. Rev. B {\bf 54}, 16470 (1996); 
S.\ Y.\ Savrasov and D.\ Y.\ Savrasov, Phys. Rev. B {\bf 54} 16487 (1996).

\bibitem{Massidda}G. Profeta, A. Continenza, and S. Massidda, Phys. Rev. B \textbf{68}, 144508 (2003). 

\bibitem{Pick-Cdop}D.\ Kasinathan, K.-W.\ Lee, W.\ E.\ Pickett,
\eprint{cond-mat/0409563} (unpublished).
%%% On Heavy Carbon Doping of MgB$_2$

\bibitem{MazErw} S. C. Erwin and I. I. Mazin,  Phys. Rev. B \textbf{68}, 132505 (2003).
 
 
\end{thebibliography}
\end{document}